# Rotating Capacitive System for Measurement of a Third Order Gravity Gradient Based on a Single-Axis Mechanical Oscillator. Part I, Theory.

Alexey V. Veryaskin[1] , Khyl J. McMahon[2] and Howard C. Golden[3]

[1]Robinson Research Institute, Victoria University of Wellington, 69 Gracefield Road, Lower Hutt 5010, New Zealand
[2]AUX Ltd, 65 Dundas Road, Inglewood WA 6052, Australia
[3]School of Physics, Mathematics and Computing, University of Western Australia, 35 Stirling Hwy, Crawley WA 6009, Australia

Corresponding author: Alexey V. Veryaskin ( alex.veryaskin@vuw.ac.nz )

## Abstract

This paper describes a single-axis mechanical oscillator, mounted on a miniature circular rotating platform that is proposed to be used for measuring a quantity proportional to the second vertical derivative of a gravity acceleration projected onto the oscillator's sensitivity axis. The test mass of the mechanical oscillator is a floating metal plate rotating with the platform. The test mass displacement along its sensitivity axis is confined to one direction only, which is also aligned with the platform's axis of rotation. The oscillator's test mass acts as a gating, floating capacitive electrode crossing two adjacent, flat capacitive electrodes positioned stationary with a fixed distance between them and separated from the test mass by a small air gap. Four such electrodes, connected in parallel, represent a primary, fixed capacitor. While crossing the stationary electrodes, the oscillator's test mass changes the primary capacitance due to electrical coupling between them. The gap between the test mass and the stationary electrodes depends on variations in the gravity acceleration component and other mechanical disturbances that may occur due to the rotating platform's motion with respect to a stationary reference coordinate frame. In turn, this results in a double triparty capacitance modulation by the platform's rotation and, as a result, the periodic variations of the gravitational acceleration caused by gravity gradients in the rotation plane. An ultra-sensitive capacitance-to-phase converter is used as an integral part of a precision RF (1-3 MHz) Mach-Zehnder interferometer, converting the capacitance variations into voltage variations at its output. The converter is based on a modified All-Pass Filter (APF) topology [1], resulting from its latest version approximately 48 V/pF of the interferometer transfer function and ~40 nV/√Hz output voltage noise. Its output phase-modulated signal concludes the analogue part of the proposed system and further signal processing is done in a FPGA software-based environment. The content of this publication represents a feasibility study aimed at identifying the practical implementation of the measurement process and the system design, which would allow for the detection of the second spatial derivative of gravity. It also aims at considering the system's potential deployment from a drone and/or other moving platforms. A parametric modulation of the desired signal due to variable electrostatic coupling between the accelerometer's rotating test mass and the stationary capacitive electrodes is also considered in this study. Part I establishes theoretical feasibility only. Design constraints, quantitative sensitivity, error budgets, and mobile-platform performance will be addressed in Part II.

## 1. Introduction

Gravity gradients serve as highly valuable sources of information across a range of strategic applications. These fields of application include oil and gas exploration, mineral exploration, the detection of subsurface voids, tunnels, and similar underground structures, heavy mass detection, passive GPS-denied navigation, and various defence-related operations [2]. Gravity gradients represent the first spatial derivatives of the gravitational acceleration vector $\vec{g}$

$$\frac{\partial}{\partial x} g_x = G_{xx},\ \frac{\partial}{\partial y} g_y = G_{yy},\ \frac{\partial}{\partial z} g_z = G_{zz}$$
$$\frac{\partial}{\partial y} g_x = \frac{\partial}{\partial x} g_y = G_{yx} = G_{xy},$$
$$\frac{\partial}{\partial z} g_x = \frac{\partial}{\partial x} g_z = G_{zx} = G_{xz}$$
$$\frac{\partial}{\partial z} g_y = \frac{\partial}{\partial y} g_z = G_{zy} = G_{yz} \qquad (1)$$

Gravity gradient components obey the following Laplacian identity

$$G_{xx} + G_{yy} + G_{zz} = \Omega^2 = {\Omega_x}^2 + {\Omega_y}^2 + {\Omega_z}^2 \quad (2)$$

where $\Omega_x, \Omega_y, \Omega_z$ the cartesian components of the angular rotation vector in the rotating coordinate frame.

The airborne deployable Falcon generation of rotating gravity gradiometers, developed by Lockheed Martin Corporation (LMC), is able to simultaneously measure the $G_{xy}$ component and the combination $G_{yy}$-$G_{xx}$ which is called the differential gravity field curvature [3]. LMC's Full Tensor Gradiometer (FTG) rotating type systems are able to measure all gravity gradient components but their stable operational time is much slower than the later Falcon's versions one [4]. Classic and Quantum free fall gravity gradiometers can measure the $G_{zz}$ component under stationary conditions [2,5], and there are proposals for measuring different components as well [6]. The distributed type of gravity gradiometers developed by Gravitec Instruments (UK, NZ, AU) group of companies in 2000-2009 [8], and later, in 2016-2023, by the University of Western Australia, can measure $G_{zx}$ and $G_{zy}$ components or their combinations [9]. A full classification of all gravity gradiometer systems and their operational principles can be found in [2].

Currently, the Falcon and FTG generations of gravity gradiometers are the only mobile types of precision gradiometric instruments capable of being used on airborne surveying platforms. However, they are high-cost, bulky, cumbersome, heavy, and have high power requirements. These types of gravity gradiometers are used on medium-sized aircrafts, helicopters, and strategic submarines but cannot be installed on drones, AUVs, UAVs, and other unmanned platforms, including submersibles. Miniaturized gravity gradiometers suitable for such carriers are seen as a strategic new direction in mobile gravity gradiometry that can revolutionize the technology and its applications [10,11]. All gravity gradiometers intended for use on mobile platforms suffer from the same problem: what is measured is not pure gravity gradients, but rather quantities containing non-inertial terms arising from the platform's angular motion. In order to cancel out such contamination, specifically designed mounting frames are used, which provide active compensation for any unwanted degree of freedom of the mobile platforms. These frames are called stable tables, and they may cost much more than the gradiometers themselves.

The system under consideration in this study aims to be compatible with deployment on manned and unmanned mobile platforms and to directly measure the signal proportional to the following quantity

$$G_{ZZZ} = \frac{\partial}{\partial Z} G_{ZZ} \quad (3)$$

where the Z-axis is assumed to be aligned with the plumb line and the platform's axis of rotation. This quantity is called the vertical spatial derivative of the $G_{ZZ}$ gravity gradient component which is almost impossible to measure directly among the other components shown in Eq.1. If the vertical gravity gradient $G_{ZZ}$ is measured in a coordinate frame experiencing rotations with respect to stationary frame of reference, then the so-called dynamic gravity gradient is one which is measured, containing the squared angular rotation frequencies as shown below

$$\widetilde{G_{ZZ}} = G_{ZZ} + {\Omega_X}^2 + {\Omega_Y}^2 \quad (4)$$

However, the angular rotation terms do not depend on spatial coordinates and this makes the quantity in Eq.(3) independent of the rotational contamination which otherwise, in practice, needs to be mitigated by using complex and expensive countermeasures. The authors of this study are not aware of any proposals for measuring the third order vertical gravity gradient except the one discussed in [12,13].

## 2. A single-axis mechanical oscillator in a rotating coordinate frame

A single-axis mechanical oscillator is mounted in a radial position $r_0$ on a disk rotating at angular rate Ω, with its sensitive axis parallel to the vertical axis Z, as depicted in Fig.1. For the oscillator test mass displacement ΔZ, velocity $V_z$ = dΔZ/dt, mass m, resonant angular frequency $\omega_0 = 2\pi f_0$ and relaxation time τ, one has

$$\frac{d^2}{dt^2}\Delta Z + \frac{2}{\tau}\frac{d}{dt}\Delta Z + \omega_0^2 \Delta Z = g_z(t) - a_z(t) + + \frac{1}{m} N_L(t) \quad (5)$$

where $a_Z(t)$ is the vertical linear acceleration of the rotating platform, $N_L(t)$ is the random Langevin noise term and $g_Z(t)$ is the projection of gravitational acceleration vector on the oscillator's sensitivity axis Z. Furthermore, we consider an example of such a mechanical oscillator as a cantilever flexor fixed at one end and having the test mass at the other, as shown in Fig. 3. However, different types of single-axis mechanical oscillators can be used in practical implementation of the system under consideration. If the oscillator's sensitivity axis is exactly aligned with the axis of rotation, then Eq.5 does not contain any non-inertial terms originating from Coriolis, Centrifugal and Euler forces which are present in any rotating coordinate frame [14]. The case when the latter does not hold is considered in the error budget analysis section in Part II of this study.

For the sake of simplicity, the vertical linear acceleration can be considered as the strongest spectral harmonic at an angular frequency ω

$$a_z(t) = F_a(\omega) a_z(\omega) Sin(\omega t) \tag{6}$$

where $F_a(\omega)$ is a spectral attenuation factor in case the rotating platform is mechanically stabilized against possible linear disturbances.

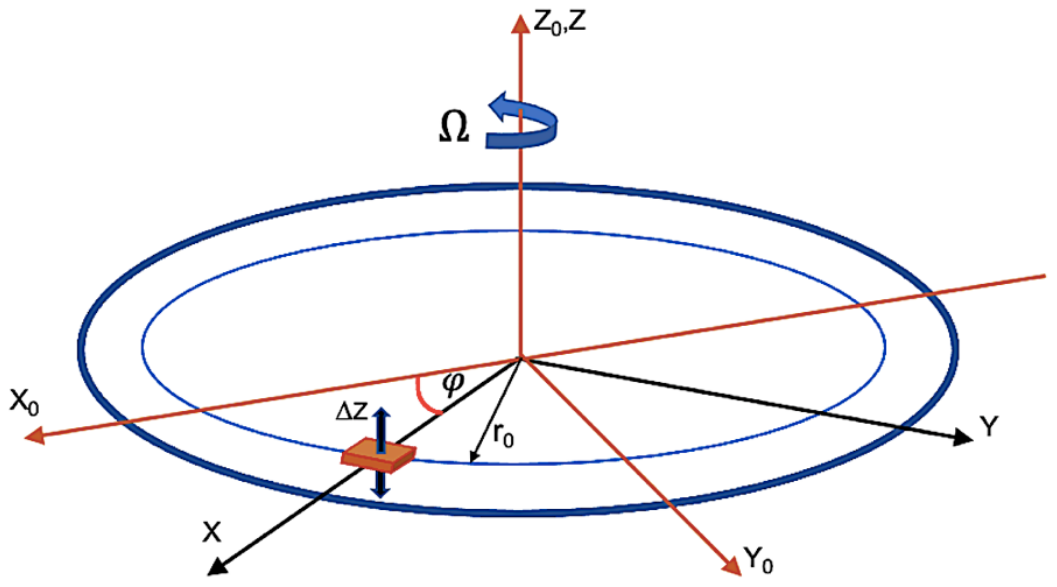

Figure 1. A single-axis mechanical oscillator is positioned along a radial direction of the rotating disk that coincides with the X-axis of the coordinate system XYZ rotating with the disk. The coordinate system ($X_0$,$Y_0$,$Z_0$) is a non-rotating local coordinate frame. The axes Z and $Z_0$ are both aligned perpendicular to the disk's plane. Here the standard right-handed Cartesian coordinate system is used where the rotation angle is positive anticlockwise.

By differentiating Eq.5 one has

$$\frac{d^2}{dt^2}V_z + \frac{2}{\tau}\frac{d}{dt}V_z + \omega_0^2 V_z = \frac{d}{dt}g_z(t) - \frac{d}{dt}a_z(t) + \\ + \frac{1}{m}\frac{d}{dt}N_L(t) \tag{7}$$

where $V_z$ is the oscillator test mass velocity and $\frac{d}{dt}g_z(t)$ is the total time derivative of the gravity acceleration component. In the case depicted in Fig.1, the latter can be expressed as [2]

$$\frac{d}{dt}g_z(t) = \frac{\partial}{\partial t}g_z(t) - r_0\Omega G_{zx} Sin(\Omega t + \phi_0) + \\ + r_0\Omega G_{zy} Cos(\Omega t + \phi_0) \tag{8}$$

where $\phi_0$, $G_{zx}$, $G_{zy}$ are consequently the initial oscillator's centre of mass angular position and the local values of the gravity gradient components with respect to the non-rotating reference frame ($X_0$,$Y_0$,$Z_0$).

Then Eq.7 can be written as follows

$$\frac{d^2}{dt^2}V_z + \frac{2}{\tau}\frac{d}{dt}V_z + \omega_0^2 V_z = \\ = -r_0\Omega G_{zx} Sin(\Omega t + \phi_0) + \\ + r_0\Omega G_{zy} Cos(\Omega t + \phi_0) - \\ - \frac{d}{dt}a_z(t) + \frac{1}{m}\frac{d}{dt}N_L(t) \tag{9}$$

The $\partial g_z(t)/\partial t$ term is excluded in Eq.9 from further consideration, since the temporal variations of gravitational acceleration, like tidal ones, does not significantly affect the measurement process. However, this term can be a calibration signal generated at the same rotational frequency Ω by a variable gravity field source.

Let $k_\Omega = \Omega/\omega_0$, $k_\omega = \omega/\omega_0$ and $Q = \omega_0\tau/2$, then the general solution of Eq.9 is as follows

$$V_z(t) = \frac{r_0\Omega}{\omega_0^2}\frac{1}{(1-k_\Omega^2)^2 + \left(\frac{k_\Omega}{Q}\right)^2} \times \\ \times \left\{ \begin{array}{l} Gzx\left[\frac{k_\Omega}{Q}Cos(\Omega t) - (1-k_\Omega^2)Sin(\Omega t)\right] + \\ + Gzy\left[\frac{k_\Omega}{Q}Sin(\Omega t) + (1-k_\Omega^2)Cos(\Omega t)\right] \end{array} \right\} - \\ - \frac{1}{\omega_0^2}\frac{\omega F_a(\omega) a_z(\omega)}{(1-k_\Omega^2)^2 + \left(\frac{k_\Omega}{Q}\right)^2} \times \\ \times \left[\frac{k_\omega}{Q}Sin(\omega t) + (1-k_\omega^2)Cos(\omega t)\right] + v_z(t) \tag{10}$$

where $v_z(t)$ is the velocity noise term [14]

$$v_z(t) = \\ = \frac{1}{m\varpi}\int_{-\infty}^{t} d\xi \frac{d}{dt}N_L(\xi) Sin[\varpi(t-\xi)] e^{-\frac{t-\xi}{\tau}} \tag{11}$$

and

$$\varpi = \sqrt{\omega_0^{\,2} - \frac{1}{\tau^2}}.$$

By taking the integral in Eq.11 by parts, one has

$$v_z(t) = \\ = \frac{d}{dt}\left\{\frac{1}{m\varpi}\int_{-\infty}^{t} d\xi\, N_L(\xi) Sin[\varpi(t-\xi)] e^{-\frac{t-\xi}{\tau}}\right\} \\ = \frac{d}{dt}\delta Z(t) \tag{12}$$

By using Eq.12 and Eq.10, one can find the general solution of Eq.5 for the oscillator mechanical displacement ΔZ(t) including the Langevin noise term

$$\Delta Z(t) = \frac{r_0}{\omega_0^2}\frac{1}{(1-k_\Omega^2)^2+\left(\frac{k_\Omega}{Q}\right)^2}\times$$
$$\times\left\{\begin{array}{l} Gzx\left[\frac{k_\Omega}{Q}Sin(\Omega t)+(1-k_\Omega^2)Cos(\Omega t)\right]+ \\ +Gzy\left[-\frac{k_\Omega}{Q}Cos(\Omega t)+(1-k_\Omega^2)Sin(\Omega t)\right]\end{array}\right\}$$
$$+\frac{1}{\omega_0^2}\frac{F_a(\omega)a_z(\omega)}{(1-k_\Omega^2)^2+\left(\frac{k_\Omega}{Q}\right)^2}\times$$
$$\times\left[-\frac{k_\omega}{Q}Cos(\omega t)+(1-k_\omega^2)Sin(\omega t)\right]+$$
$$+\delta Z(t) \qquad (13)$$

where

$$\delta Z(t) =$$
$$= \frac{1}{m\varpi}\int_{-\infty}^{t} d\xi\, N_L(\xi)Sin[\varpi(t-\xi)]e^{-\frac{t-\xi}{\tau}} \qquad (14)$$

is the corresponding oscillator test mass response to the Langevin noise source. The latter obeys the standard auto-correlation rule

$$\langle N_L(t)N_L(t')\rangle = \frac{4mk_BT}{\tau}\delta(t-t') \qquad (15)$$

where $k_B$ is the Boltzmann constant and T is the ambient temperature.

## 3. Capacitive modulation geometrical model

Consider four stationary capacitors $C_0$ formed by pairs of adjacent flat metal strips, which are connected in parallel to one sensing rail and the companion rail is grounded as shown in Fig.2. The variable gating capacitors C(t) are formed by an effective overlap area $A_{eff}(t)$ between the oscillator test mass and the stationary strips separated by gap $d_0+\Delta Z(t)$ while the former is crossing the latter along a circle, as depicted in Fig.3. The gap depends on the oscillator's test mass mechanical displacement ΔZ(t).

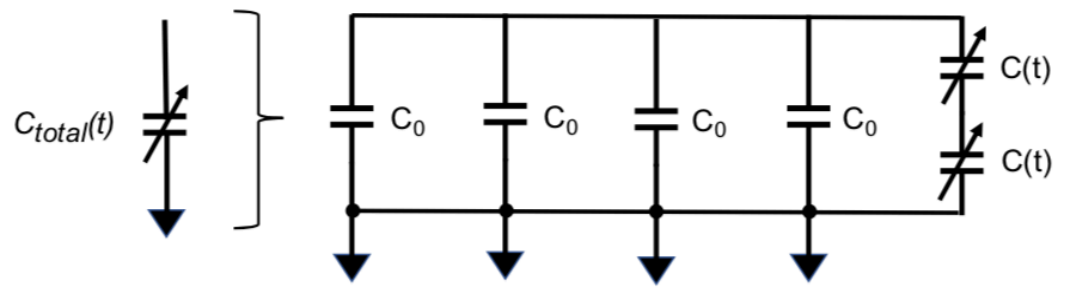

Figure 2. Equivalent circuit of the capacitive sensing system. Four fixed capacitors $C_0$ are connected in parallel between the common sensing rail and ground. The electrically floating rotating test-mass electrode forms two variable coupling capacitors C(t) in series, contributing C(t)/2 to the total capacitance when their values are treated equal. The case when this does not hold will be considered in Part II.

The equivalent total capacitance is as follows

$$C_{total}(t) =$$
$$= 4C_0 + \frac{1}{2}C(t) = 4C_0 + \frac{1}{2}\frac{\varepsilon_0\varepsilon A_{eff}(t)}{d_0+\Delta Z(t)} \qquad (16)$$

where $\varepsilon_0$ is the free air dielectric constant and $\varepsilon$ is the relative dielectric constant which is greater than one if the capacitive plates are coated by thin layers of advanced dielectric materials as is widely used in MEMS capacitive sensors [16].

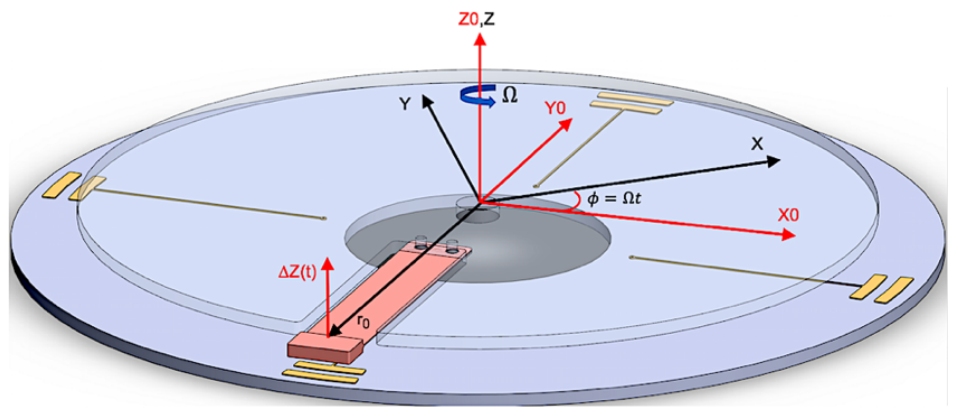

Figure 3. Three-dimensional representation of the rotating capacitive gradiometer. A single-axis oscillator is mounted at radial offset $r_0$ on a platform rotating at angular velocity Ω. Its floating test-mass electrode undergoes axial displacement ΔZ(t) while sequentially traversing four stationary electrode pairs. The overlap-dependent coupling and displacement-dependent gap jointly modulate the measured capacitance.

Without loss of generality, let the rotating electrode occupy the radial interval $r_1 \le r \le r_2$ with angular width β, and let each stationary capacitive electrode have an angular width α (see Fig.4). Both widths are expressed in radians. Fig.3 above shows a particular case when α=β. The rotor angle is

$$\emptyset(t) = \Omega t + \emptyset_0$$

Let the four stationary electrode pairs be centred at

$$\theta_n = \frac{n\pi}{2}, \qquad n = 0,1,2,3\ldots \qquad (18)$$

Further, define the radial area coefficient

$$K_A = \frac{r_2^2 - r_1^2}{2} \qquad (19)$$

For an angular overlap of γ radians, the geometrical overlap area between the rotating and stationary electrodes is exactly

$$A_{eff}(t) = K_a\gamma(t) \qquad (20)$$

The overlap function $\gamma(t)$ is the circular cross-correlation of two angular windows. For the electrodes' widths satisfying α + β ≤ 2π, one can define the shortest signed angular separation

$$\delta_n(t) = wrap_{(-\pi,\pi)}(\Omega t + \emptyset_0 - \theta_n)$$

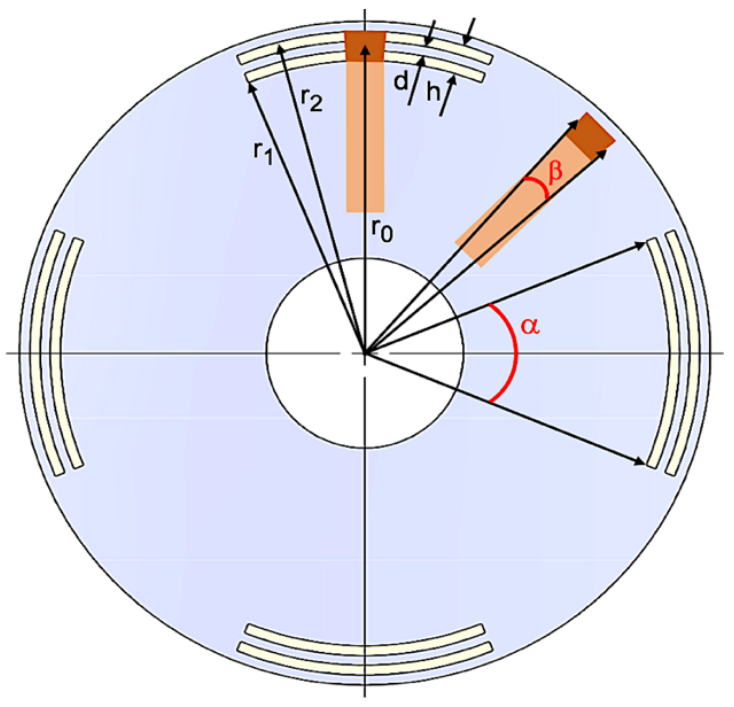

Figure 4. Plan view of the electrode geometry. The rotating electrode occupies $r_1 \leq r \leq r_2$ and angular width β, while each of the four stationary electrode pairs has angular width α and is centred at an azimuth separated by π/2 from its neighbours. The nominal oscillator radius is $r_0$. The diagram also indicates electrode clearances and the gap geometry. Fourfold symmetry suppresses all geometrical overlap harmonics except integer multiples of 4Ω in the ideal device.

Then the angular overlap is then as follows

$$\gamma_n(t) = $$
$$= \begin{cases} \min(\alpha,\beta), & |\delta_n| \leq \dfrac{|\alpha-\beta|}{2} \\ \dfrac{\alpha-\beta}{2} - |\delta_n|, & \dfrac{|\alpha-\beta|}{2} < |\delta_n| < \dfrac{\alpha+\beta}{2} \\ 0, & |\delta_n| \geq \dfrac{\alpha+\beta}{2} \end{cases}$$

and

$$A_n(t) = K_A \gamma_n(t) \tag{21}$$

The overlap rises linearly as the rotating electrode enters a stationary sector, remains constant if α >> β, and decreases linearly as the rotating electrode leaves it.

The effective geometrical overlap is as follows

$$A_{eff}(t) = \sum_{0}^{3} A_n\left(\Omega t + \emptyset_0 - \frac{n\pi}{2}\right)$$

where $A_n(t)$ denotes the single-pair overlap function defined in Eq.21.

For exactly identical pairs one has

$$A_{eff}\left(t + \frac{\pi}{2\Omega}\right) = A_{eff}(t)$$

Consequently, although the rotor completes one revolution in 2π/Ω time interval, the effective overlap repeats for times per revolution. So, its fundamental modulation frequency is 4Ω.

The Fourier series coefficients can be derived analytically without approximating the general overlap trapezoidal waveform. Let $A_0(\theta)$ be the overlap area of one stationary electrode pair centred at zero angle. Its Fourier expansion is

$$A_0(\theta) = c_0 + 2\sum_{k=1}^{\infty} c_k\, Cos(k\theta)$$

where coefficients are

$$c_m = \frac{1}{2\pi}\int_{-\pi}^{\pi} A_0(\theta) e^{-jk\theta} d\theta$$

Because the overlap function is the circular convolution of the rotating and stationary angular windows, its Fourier coefficients are products of their respective window coefficients

$$c_k = \frac{2K_A}{\pi k^2} Sin\left(\frac{k\alpha}{2}\right) Sin\left(\frac{k\beta}{2}\right), \quad k \neq 0$$

The dc coefficient is as follows

$$c_0 = \frac{K_A \alpha\beta}{2\pi}$$

The complete Fourier series of one electrode pair is

$$A_{eff}(\theta) = $$
$$= \frac{K_A\alpha\beta}{2\pi} + \frac{4K_A}{\pi}\sum_{k=1}^{\infty} \frac{Sin\left(\frac{k\alpha}{2}\right) Sin\left(\frac{k\beta}{2}\right)}{k^2} Cos(k\theta)$$

This result applies to the periodic angular-overlap model, including the wraparound at the end of each revolution.

The four shifted by π/2 angle electrode pairs, connected in parallel, produce the harmonic sum

$$\sum_{n=0}^{3} e^{-jkn\pi/2}$$

This sum is zero unless k is an integer multiple of four. Replacing k = 4n ( n = 1,2,3…∞ ), one finds the complete analytical Fourier expansion for the assumed ideal geometry

$$A_{eff}(t) = \frac{2K_A\alpha\beta}{\pi} +$$
$$+\frac{K_A}{\pi}\sum_{n=1}^{\infty}\frac{Sin(2n\alpha)Sin(2n\beta)}{n^2}\times$$
$$\times Cos(4n\Omega t + 4n\emptyset_0) \quad (22)$$

By substituting Eq.22 into Eq.16 and assuming that $\Delta Z(t) \ll d_0$, one has

$$C_{total}(t) = 4C_0 + \frac{\varepsilon_0\varepsilon}{d_0}\frac{K_A\alpha\beta}{\pi} + \frac{1}{2}\frac{\varepsilon_0\varepsilon}{d_0}\frac{K_A}{\pi}\times$$
$$\times\sum_{n=1}^{\infty}\frac{Sin(2n\alpha)Sin(2n\beta)}{n^2}Cos(4n\Omega t + 4n\emptyset_0) -$$
$$-\frac{\varepsilon_0\varepsilon}{d_0^2}\frac{K_A\alpha\beta}{\pi}\Delta Z(t) - \frac{1}{2}\frac{\varepsilon_0\varepsilon}{d_0^2}\frac{K_A}{\pi}\Delta Z\times$$
$$\times\sum_{n=1}^{\infty}\frac{Sin(2n\alpha)Sin(2n\beta)}{n^2}Cos(4n\Omega t + 4n\emptyset_0) \quad (23)$$

where $\Delta Z(t)$ is presented in Eq.13. One can see that the leading capacitance modulation harmonics in Eq,23 are ω, Ω, 4Ω, and the corresponding intermodulation sidebands are 3Ω, $4\Omega\pm\omega$, 5Ω. The lowest intermodulation harmonic 3Ω provides the basis for the synchronous demodulation and further signal processing as will be shown in Part II of this study. It is worth noting that the 3Ω component is a readout frequency conversion product and not, by itself, is associated with a mechanical force.

## 4. Parametric modulation of the oscillator response

In the practical measurement system, which will be presented in Part II of this study, the total sensing capacitor is an integral part of the capacitive-to-phase converter [1] pumped with a RF carrier. The latter produces a mean-square voltage $<V^2>$ across the capacitor's electrodes and this results in the variable electrostatic force between them. From Eq.16 one has for $\Delta Z(t) \ll d_0$ [17]

$$F_Z^e(t) = \frac{1}{2}\langle V^2\rangle\frac{\partial C_{total}}{\partial \Delta Z} = -\frac{1}{4}\langle V^2\rangle\frac{\varepsilon_0\varepsilon A_{eff}(t)}{[d_0+\Delta Z(t)]^2} \cong$$

$$\cong F_{Z,0}^e(t) + k_e(t)\Delta Z(t) \quad (24)$$

where $A_{eff}(t)$ is presented in Eq.22, and

$$F_{Z,0}^e(t) = -\frac{1}{4}\langle V^2\rangle\frac{\varepsilon_0\varepsilon A_{eff}(t)}{d_0^2} \quad (25)$$

The time-dependent electrostatic stiffness is

$$k_e(t) = \frac{1}{2}\langle V^2\rangle\frac{\varepsilon_0\varepsilon A_{eff}(t)}{d_0^3} \quad (26)$$

Eq.5 is now replaced by the following

$$\frac{d^2}{dt^2}\Delta Z + \frac{2}{\tau}\frac{d}{dt}\Delta Z + \omega_{eff}^2\big(1 - h(t)\big)\Delta Z =$$
$$= g_z(t) - a_z(t) + \frac{1}{m}N_L(t) -$$
$$-\frac{1}{4}\langle V^2\rangle\frac{\varepsilon_0\varepsilon A_{eff}(t)}{md_0^2} \quad (27)$$

where the dimensionless parametric modulation factor h(t) is as follows

$$h(t) = \frac{1}{2}\langle V^2\rangle\frac{\varepsilon_0\varepsilon K_A}{\pi m\omega_{eff}^2 d_0^3}\times$$
$$\times\sum_{n=1}^{\infty}\frac{Sin(2n\alpha)Sin(2n\beta)}{n^2}Cos(4n\Omega t + 4n\emptyset_0) \quad (28)$$

and

$$\omega_{eff}^2 = \omega_0^2 - \frac{1}{2}\langle V^2\rangle\frac{\varepsilon_0\varepsilon K_A\alpha\beta}{\pi m d_0^3} \quad (29)$$

The effective mechanical resonant frequency results from the electrostatic spring softening effect as the effective overlap area function $A_{eff}(t)$ in Eq.26 contains a constant term.

Eq.27 is a non-homogeneous Hill-type equation similar to that of discussed in [18]. A theoretical conceptual design for a spaceborne radially vibrating, rotating gravitational gradient sensor was introduced by Chobotov in 1968. It was based on a single mechanical oscillator positioned on a rotating disk with its sensitivity axis aligned with radial direction, i.e. aligned with the surface of the disk under rotation. This proposal has not resulted in any further developments but it is an interesting example of diverging from the traditional gravity gradiometry, as introduced here as well.

If the parametric modulation factor satisfies the condition $h(t) \ll 1$, which will be provided in the practical case, then the solution of Eq. 27 can be represented as

$$\Delta Z_{np}(t) + \Delta Z_p(t), \quad \Delta Z_p(t) \ll \Delta Z_{np}(t) \quad (30)$$

where $\Delta Z_{np}(t)$ is the non-parametric oscillator response represented in Eq.13 where $\omega_0$ is replaced by $\omega_{eff}$, and $\Delta Z_p(t)$ is the oscillator response to the parametric electrostatic modulation

$$\frac{d^2}{dt^2}\Delta Z_p + \frac{2}{\tau}\frac{d}{dt}\Delta Z_p + \omega_{eff}^2\Delta Z_p =$$
$$= -\frac{1}{4}\langle V^2\rangle\frac{\varepsilon_0\varepsilon A_{eff}(t)}{md_0^2} + \omega_{eff}^2 h(t)\Delta Z_{np}(t) \quad (31)$$

In order to further simplify the treatment of the mechanical displacement of linearly and parametrically driven oscillator's test mass, let the angle size of the stationary capacitive electrodes α be equal exactly to π/4, or 45 degree. Then one has

$$Sin(2n\alpha) = \frac{1 + (-1)^{n+1}}{2}$$

which is equal to that for n = 1, 3, 5…, and equal to zero otherwise. This means that in Eq.22 and Eq.28, the effective overlap area modulation frequencies are 4Ω, 8Ω, 12Ω,… On the other hand the leading intermodulation sidebands in Eq.23 and Eq.31 are 3Ω, $4\Omega \pm \omega$, 5Ω, 7Ω,… In order to Eq.30 hold true, none of the above modulation frequencies should coincide with the liner and parametric resonant conditions, which means

$$\frac{\omega_{eff}}{3} < \Omega < \omega_{eff} \qquad (32)$$

and $\Omega \cong \frac{3\omega_{eff}}{4}$ is an obvious choice.

One can show that the lowest driving frequency 3Ω in Eq.31 under the above conditions is more than twice that of the resonant frequency $\omega_{eff}$ so all the higher harmonics can be ignored due to the high roll-off rate ~ 12 dB per octave. One can prove that a low enough RMS voltage across the APF sensing capacitor, the ΔZp(t) correction in Eq.30 can be reduced to a second order one, and that a trade-off can be established between the required sensitivity and the sensing capacitor design features. As was pointed out in Section 3 above, the 3Ω harmonic in $C_{total}(t)$ forms the basis for the IF demodulation of the terms proportional to $G_{ZX}$ and $G_{ZY}$ gravity gradient components. It is also worth noting that the $4\Omega \pm \omega$ harmonic content in $C_{total}(t)$ can be used to demodulate the term proportional to the parasitic linear acceleration $a_z(t)$ in Eq.6 and use this channel to actively remove its potential leakage into the desired gravity gradient signal.

## 5. Analogue Front-End Interferometric Readout and Capacitance-to-Phase Conversion

The proposed instrument employs a coherent radio-frequency (RF) Mach–Zehnder-type interferometric readout. An oscillator disciplined by a precision oven-controlled crystal oscillator (OCXO) provides the reference for a direct digital synthesiser (DDS). A power splitter generates two mutually coherent signals. The sensing arm contains a modified resonant all-pass filter (APF), here designated $K_{APF}$, whose phase response is controlled by the total sensing capacitance $C_{total}(t)$ in Eq.23 . The other arm serves as the RF reference. The arms are recombined so that their nominal carrier components cancel, leaving a residual signal sensitive to changes in the sensing-arm complex transfer function. The sensing capacitance is the single electrical quantity presented by the four stationary electrode positions connected to a common sensing rail and the companion rail is grounded. A rotating, electrically floating gating electrode modifies the equivalent capacitance. Thus the RF front end reads one time-dependent capacitance, $C_{total}(t)$, rather than four independent channels as depicted in Fig.5 below.

Let $\omega_c$ denote the RF carrier angular frequency. Let us write the $K_{APF}$ transfer function as a complex quantity H(ω,C) where a(ω,C) is its voltage-amplitude transmission and φ is its phase

$$H(\omega,C) = a(\omega,C) \exp[i\varphi(\omega,C)] \qquad (33)$$

For a capacitance perturbation δC(t) about the nominal value $C_0$, at fixed carrier frequency $\omega_c$, the leading phase response is

$$\delta\varphi(t) \simeq K_{APF}\, \delta C(t), \quad K_{APF} \equiv (\partial\varphi/\partial C)\,\big|\,(\omega_c,C_0) \qquad (34)$$

Likewise, the amplitude perturbation is $\delta a(t) \simeq K_a\, \delta C(t)$, where $K_a = (\partial a/\partial C)$ evaluated at the same operating point. The full differential response is therefore

$$\delta H/H_0 \simeq [K_a/a_0 + i\, K_{APF}]\, \delta C(t) \qquad (35)$$

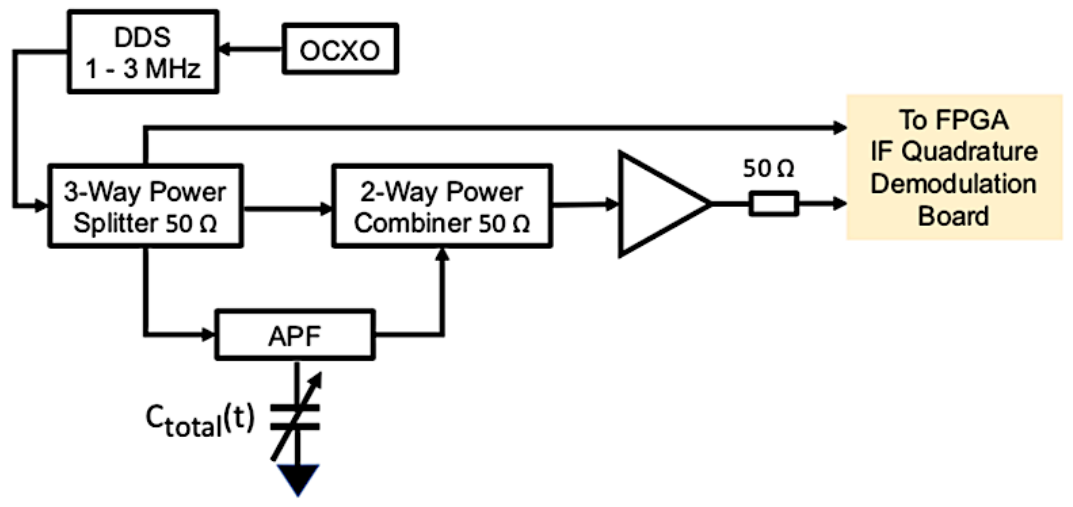

Figure 5. A simplified block diagram of the analogue front-end interferometric readout showing the APF capacitive-to-phase interface included into a measurement arm of the March-Zehnder RF interferometer. The output of the latter goes straight into FPGA-based IF demodulation stage for further signal processing.

The phase coefficient $K_{APF}$ has units rad/pF and its sign depends on the adopted phase convention. It must be distinguished from the sensitivity of the complete interferometer, which also depends on carrier amplitude, arm balance, combiner response and post-combiner gain. Using complex RF phasors, let the sensing arm signal at the combiner be $E_s = e_{s0} \exp(i\varphi_s)$ and the reference-arm signal be $Er = e_{r0}\exp(i\varphi_r)$. The combiner polarity is absorbed into the definition of the reference phasor.

The residual carrier is

$$E_{INT} = E_s + E_r \qquad (36)$$

At the intended cancellation point, $E_{s0} \approx E_{r0}$ and $\theta = \varphi_s - \varphi_r \approx \pi \pmod{2\pi}$. Near this operating point, let us introduce an amplitude mismatch $\delta a/a_0$ and a differential phase error $\delta\theta$, with $E_s = E_0(1+\delta a)\exp[i(\pi+\theta)]$ and $E_r = E_0$.

To first order,

$$E_{INT} \simeq -E_0(\delta a/a_0 + i\,\delta\theta)$$
$$|E_{INT}| \simeq E_0 \sqrt{(\delta a/a_0)^2 + \delta\theta^2} \quad (37)$$

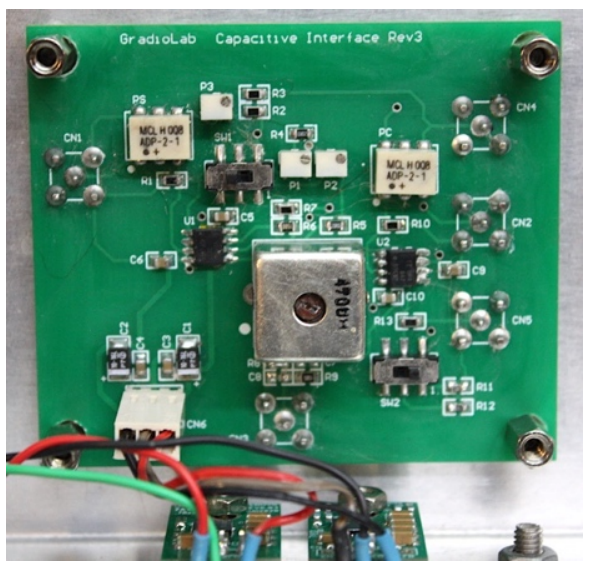

Figure 6. An image of the latest version of experimentally implemented capacitance-to-phase interface along with a RF Mach-Zehnder architecture.

The result illustrates an important qualification: the magnitude of the nulled RF signal is not a globally signed linear phase detector. At perfect balance it varies as $|\theta|$ and a signed, linear response requires a defined phase-sensitive detection quadrature, an intentional calibrated offset, or another suitable coherent demodulation method. The RF residual is amplified before ADC acquisition, while coherent phase-sensitive processing can be implemented digitally.

For a selected, calibrated phase quadrature, let $Q_\phi$ represent the net conversion from differential phase (radians) to the demodulated voltage, including the RF chain and digital scaling. The small-signal voltage response is

$$\delta V_Q(t) \simeq Q_\phi\, \delta\varphi(t) = K_{APF}\, Q_\phi\, \delta C(t) \quad (38)$$

Eq.(38) defines the effective capacitance-to-voltage conversion coefficient $K_{CV} = K_{APF} Q_\phi$ of the complete measurement channel. It should be established experimentally, rather than inferred from $K_{APF}$ alone. An amplitude-sensitive quadrature may be used simultaneously to monitor imbalance and diagnose parasitic amplitude modulation. The measured $K_{APF}$ phase response, as depicted in Fig.7 below, exhibits a steep transition in the vicinity of its resonant operating region. The carrier can be placed at a point where the sensing-arm phase is nominally opposite to that of the reference arm, avoiding a separate conventional RF phase shifter when the remaining fixed path phase is compatible with cancellation. Fine DDS frequency control provides a practical phase-balancing degree of freedom. The carrier frequency should not be chosen solely from the largest slope $|\partial\varphi/\partial C|$. The operating point must also provide adequate arm-amplitude matching, stable residual carrier, acceptable response linearity over the expected capacitance excursion, and tolerable conversion of carrier-frequency fluctuations into differential phase noise. Frequency tuning and amplitude balancing are therefore coupled calibration tasks.

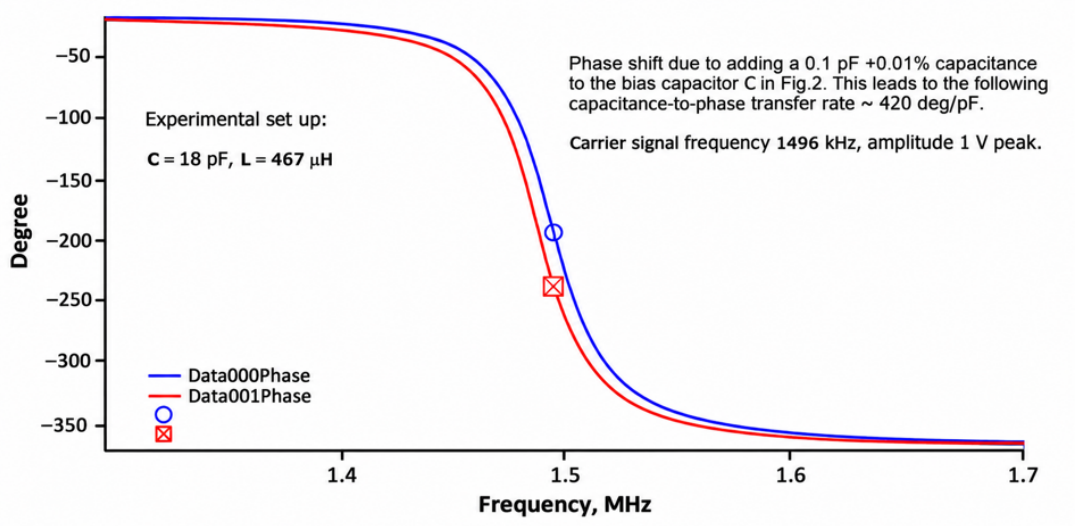

Figure 7. Experimental phase-versus-frequency response of the $K_{APF}$ under two sensing-capacitance conditions. The steep transition illustrates capacitance-to-phase conversion and the possibility of selecting a carrier frequency near the phase-opposition condition.

A perturbation in the synthesiser frequency produces differential phase noise whenever the sensing and reference arms have unequal phase dispersion. For small fluctuations $\delta\omega(t)$, the phase error contains the term

$$\delta\theta(t) \simeq K_{APF}\, \delta C(t) + D_\omega\, \delta\omega(t) \quad (39)$$

$$D_\omega \equiv \partial\theta / \partial\omega \,\big|_{\omega_c} \quad (40)$$

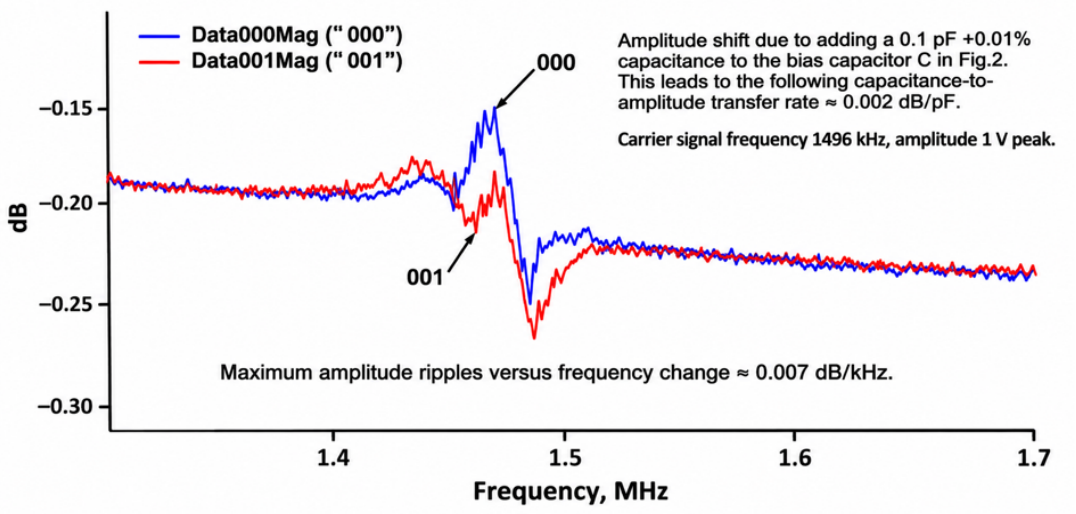

Figure 8. Experimental amplitude-versus-frequency response under the corresponding capacitance conditions.

A common OCXO/DDS source suppresses phase noise that is strictly common to both paths, but it does not automatically cancel differential phase noise caused by unequal group delay, phase dispersion, or environmental changes. The relevant source-noise spectrum must be evaluated at offsets corresponding to the recovered modulation frequencies and with the actual interferometer transfer function. An ideal all-pass network has constant magnitude, but a practical resonant $K_{APF}$ includes loss, finite component quality factors and parasitic coupling. Despite the very low experimental ripple level, shown in Fig.8, amplitude fluctuations can leak into the selected phase quadrature through imperfect I/Q orthogonality, imbalance or amplifier nonlinearity. These effects will be discussed in the quantitative noise and error budget analysis in Part II of this study.

A practical calibration sequence can use a controlled capacitance perturbation, or a traceable equivalent RF test stimulus, to determine $K_{CV}$ at the chosen operating point. A coherent digital receiver can estimate the complex residual phasor, separate its in-phase and quadrature components, and adjust the DDS carrier frequency to minimise the phase mismatch. Amplitude balance may be achieved through a stable electronically controlled gain or attenuation element if required. Alternatively, its measured residual can be monitored and corrected within a validated linear range. A periodic, rather than continuously intrusive, calibration is preferable if long-term stability permits. The low-noise RF amplifier drives the ADC. Secondary demodulation and subsequent numerical differentiation are performed in the FPGA/digital domain, not by the analogue $K_{APF}$ itself. The known deterministic overlap modulation is removed or rejected by synchronous processing before interpreting the remaining signal as gravity gradients related test-mass motion.

## 6. The estimator of the vertical spatial derivative of the $G_{ZZ}$ gravity gradient component

As will be disclosed in the Part II of this study, the proposed $G_{ZZZ}$ estimator is characterised by a triple modulation-demodulation frequency scheme because the RF carrier phase is modulated at both the rotation frequency and its higher-order harmonics (primarily the third and fourth harmonics) as the rotating gating electrode successively traverses the four stationary electrodes. By sampling the capacitive variation proportional signal at four distinct, 90-degree spaced angular positions, as depicted in Fig.4, the estimator is physically configured to measure the local gravity gradient components $G_{ZX}$ and $G_{ZY}$ at four orthogonal angular coordinate points during a single revolution. This spatial sampling is equivalent to establishing a local measurement time grid synchronised with the rotation frequency Ω

$$t_{n,k} = \frac{2\pi k + \frac{\pi n}{2}}{\Omega} \qquad (41)$$

where $k$ = 0,1,2,… is the integer revolution number and $n$=1,2,3,… represents the specific electrode index corresponding to the locations of stationary electrodes. This orthogonal time sampling allows the estimator to calculate a linear combination of the gravity gradient components $G_{ZX}$ and $G_{ZY}$ proportional to the vertical derivative of the vertical gravity gradient component $G_{ZZ}$ according to Eq.2. One can show, this vertical derivative can be mathematically approximated by a finite difference equation [12]

$$\begin{aligned} &G_{ZX}(t_{n,k}) - G_{ZX}(t_{n+2,k}) + \\ &+G_{ZY}(t_{n+1,k}) - G_{ZY}(t_{n+3,k}) \cong \\ &\cong -2r_0(G_{ZXX} + G_{ZYY}) = 2r_0 G_{ZZZ} \end{aligned} \qquad (42)$$

To perform this calculation in real-time, the measurement system can use a precision optical encoder providing a high-resolution timing trigger, allowing the downstream processing electronics to sample the measurement signal precisely at the four sequential alignment instants $t_{n,k}$, $t_{n+1,k}$, $t_{n+2,k}$ and $t_{n+3,k}$. Also, it is assumed that the time interval $\Delta t_k = 2\pi/\Omega$ is much smaller compared to potentially present time variations of the rotation angular frequency Ω, which can be treated as a constant quantity during a single revolution. This assumption, of course, depends on choosing the optimum mechanical oscillator design.

## Discussions

The proposed rotating capacitive architecture provides a theoretical route to the third-order gravity gradient $G_{ZZZ}$ using one mechanically sensitive axis, four stationary capacitive electrode pairs and a rotating floating gate. Its defining feature is that information about the spatial derivatives of $G_{ZX}$ and $G_{ZY}$ is obtained from measurements at orthogonal locations on a circular path. These derivatives are related to $G_{ZZZ}$ by the differentiated Laplace identity. The method therefore differs conceptually from instruments that directly compare vertically separated accelerometers to determine $G_{ZZ}$. The fourfold symmetry of the electrode configuration provides a useful frequency-domain structure. The leading geometrical overlap modulation occurs at 4Ω, and its product with the oscillator response at Ω generates a 3Ω electrical component. For high enough Ω this permits synchronous signal extraction away from the lowest-frequency readout band where most of linear and angular disturbances may occur. A principal potential benefit of the proposed rotating capacitive system is suppression of certain inertial terms. For a rigid rotor and ideal axial alignment, the dominant horizontal centrifugal and Coriolis terms have no direct axial projection, and spatial differentiation eliminates position-independent angular-velocity-squared contributions.

The interferometer architecture combines high-slope resonant phase transduction with coherent carrier suppression. The key physical advantage is that a small capacitance perturbation can be expressed as a measurable differential RF phase variation, and ultimately as a calibrated voltage signal suitable for digital processing. The existing hardware and historical phase/amplitude measurements support the feasibility of the conversion principle, while not by themselves establishing the ultimate gravity-gradient resolution. The cancellation point, frequency-to-phase coupling, residual amplitude response, coherent detection quadrature and stability of the signal measurement and processing chain are essential system-level considerations. The practically measured values, together with amplifier and ADC noise, calibration uncertainty and practical operating parameters, are reserved for Part II. This separation keeps the present chapter focused on the readout principle and its connection to the instrument's theoretical measurement chain.

The present analysis is intentionally restricted to the measurement principle. It does not establish a numerical $G_{ZZZ}$ sensitivity or demonstrate operation on a drone, AUV or other moving platform. Part II will address the detailed demodulation and reconstruction chain, oscillator Brownian noise, capacitive-to-phase and voltage noise, ADC and digital-processing contributions, finite measurement-time variance, rotation-frequency optimization and systematic motion-induced error budget. Experimental validation will ultimately be required to determine whether the theoretical benefits translate into a practical instrument. In summary, the rotating capacitive system offers a compact and physically distinct approach to higher-order gravity gradiometry. The central theoretical results are the exact fourfold overlap modulation, its electrical sideband structure, and the spatial finite-difference relation to $G_{ZZZ}$ . Their practical realization depends on controlled electromechanical coupling, a calibrated reconstruction of $G_{ZX}$ and $G_{ZY}$, and rigorous treatment of sequential sampling and platform dynamics.

The measurement of the third-order vertical gravity-gradient component $G_{ZZZ}$ may be particularly valuable in applications requiring discrimination of localized subsurface density contrasts. Potential targets include shallow tunnels, cavities, buried infrastructure, and other compact structures whose gravity fields exhibit appreciable spatial curvature. Because higher-order derivatives emphasize shorter spatial scales, the observable may help distinguish relatively localized anomalies from slowly varying regional gravity backgrounds. This potential advantage is accompanied by more rapid attenuation of the signal with increasing source distance; consequently, shallow targets and close-range surveys are likely to offer the most favorable conditions.

Further plausible applications include near-surface geological and engineering investigations, mineral exploration, and selected geothermal surveys where faults, contacts, or other density boundaries produce sufficiently strong and spatially localized signatures. In these settings, $G_{ZZZ}$ would most naturally complement conventional gravity and second-order gravity-gradient observations rather than replace them. The practical value of the additional spatial information will depend on target geometry and depth, survey altitude, geological background variability, instrument sensitivity, and the ability to control or correct platform-motion disturbances.

**Acknowledgements**

One of the authors (Veryaskin) would like to thank Dan DiFrancesko of Niagara Gravity Consulting LLC for his valuable comments and a reference provided.